\documentclass[fleqn,twoside]{article}
\usepackage{espcrc2}

\usepackage{graphicx}
\usepackage[figuresright]{rotating}

\newcommand{\be}{\begin{equation}}
\newcommand{\bea}{\begin{eqnarray}}
\newcommand{\ee}{\end{equation}}
\newcommand{\eea}{\end{eqnarray}}
\def\chic#1{{\scriptscriptstyle #1}}

\newcommand{\AmS}{{\protect\the\textfont2
  A\kern-.1667em\lower.5ex\hbox{M}\kern-.125emS}}

\title{On the definition and observability of the 
       neutrino charge radius.}

\author{J. Papavassiliou\address[MCSD]{
Departamento de F\'\i sica Te\'orica and IFIC, 
Universidad de Valencia-CSIC,\\
E-46100, Burjassot, Valencia, Spain},
         J. Bernab\'eu\addressmark,
        and
         J. Vidal\addressmark}

\begin{document}

\begin{abstract}
We present a brief summary of recent results concerning the 
unambiguous definition and 
experimental extraction 
of the gauge-invariant and 
process-independent neutrino charge radius.

\vspace{1pc}
\end{abstract}

\maketitle

\section{Introduction}

The neutrino electromagnetic form-factor and the 
neutrino charge radius (NCR) have constituted an important   
theoretical puzzle for over than three decades.
Since the dawn of the Standard Model (SM) it had 
been pointed out that 
 one-loop  radiative  corrections 
to the scattering amplitude of a charged particle 
(such as the electron) with 
a neutrino 
would generate 
an   effective  photon-neutrino
vertex ${\Gamma}^{\mu}_{A \nu \bar{\nu}}$, which, in turn,
would give rise  to   a   non-zero   
NCR \cite{Bernstein:jp}.
Traditionally (and, of course, rather heuristically) 
the NCR has been interpreted as 
 a measure of the ``size'' of the neutrino $\nu_i$ when probed 
electromagnetically, owing to  
its     
classical definition  
(in the static limit) as the second moment 
of the  spatial neutrino charge density $\rho_{\nu}(\bf{r})$, i.e. 
$\big <r^2_{\nu}\,  \big> = e^{-1} \int d {\bf{r}} r^2 \rho_{\nu}({\bf{r}})$. 
Even though in the SM the one-loop
computation of the 
{\it entire} $S$-matrix element 
describing the  electron-neutrino scattering  
is conceptually straighforward, the identification
of a {\it sub-amplitude},  
which would serve as the  effective 
${\Gamma}^{\mu}_{A \nu \bar{\nu}}$
has
been   faced   with  serious   complications,   associated  with   the
simultaneous   reconciliation   of   crucial  requirements   such   as
gauge-invariance,       finiteness,       and      target-independence
(for various references see \cite{Bernabeu:2000hf}). 
The crux of the problem is that  
since in non-Abelian
gauge   theories individual   
off-shell  Green's functions are in general unphysical, the
definition of quantities familiar from scalar theories or QED, such as
effective charges and form-factors, is in general problematic.
 In the
case of the NCR, various  attempts to define  its SM value through the
one-loop  $\gamma\nu\nu$  vertex   calculated  in the   renormalizable
($R_{\xi}$) gauges   reveal  that  the   corresponding electromagnetic
form-factor depends  explicitly  on  
the  gauge-fixing parameter $\xi$
   in a
prohibiting  way.  
In particular, even though  in  the static limit of
zero momentum transfer,  $q^2 \to 0$,  the form-factor 
becomes independent of  $\xi$,  its first  derivative with respect  to
$q^2$, which corresponds to the definition  of the NCR, 
continues to depend on it. 
Similar (and some times worse) problems occur in the context of other
gauges (e.g. unitary gauge).
These complications have obscured the 
entire concept of an NCR, and have casted serious doubts 
on whether it can be regarded as a genuine physical observable.

In the last two years significant progress has been 
accomplished in our understanding of the NCR.
To begin with, 
the field-theoretcial difficulties mentioned above
have    been    conclusively   settled    in
\cite{Bernabeu:2000hf},
by resorting to the well-defined electroweak gauge-invariant
separation of physical  amplitudes into effective self-energy, vertex
and  box  sub-amplitudes,   implemented  by the  pinch  technique
formalism \cite{Cornwall:1982zr}.
In addition, in a very recent work 
\cite{Bernabeu:2002nw},
the observable nature of the
NCR was established. In particular it was   
shown that the probe-independent  
charge radius of the neutrino  
may be extracted from experiment, at least in principle.
This was accomplished by expressing a set of  
experimental $\nu_{\mu}-e$ cross-sections in terms of 
the NCR and two additional  
gauge- and renormalization-group-invariant quantities,
corresponding to the electroweak effective charge 
and mixing angle. 
In what follows we will present a brief summary of these 
new developments. 

\section{On the PT definition of the NCR}

The  PT is  a
diagrammatic  method which exploits  the  underlying  symmetries 
encoded  in  a  {\it  physical} amplitude  such  as  an  $S$-matrix  element,
in  order  to  construct effective   Green's   functions    with   special  
properties.    The aforementioned symmetries,  even though they are  always
present, they are  usually concealed by  the gauge-fixing  procedure.  The  PT
makes them  manifest by means  of a  fixed algorithm,  which does  {\it not}
depend on  the gauge-fixing scheme  one uses in order  to quantize the theory,
{\it i.e.} regardless of the  set of Feynman rules used when writing down  the 
$S$-matrix  element.   
In particular, the  PT  uses  the elementary  Ward
identities 
triggered  by the  longitudinal  momenta appearing  inside  Feynman  diagrams  
in  order  to  enforce massive cancellations. The realization of these
cancellations  mixes non-trivially contributions stemming from diagrams of
different  kinematic nature (propagators, vertices, boxes). Thus,      a  given
physical  amplitude  is reorganized into sub-amplitudes,   which  have   the 
same   kinematic   properties  as conventional $n$-point functions and, in 
addition, are  endowed with  desirable physical  properties.  Most importantly,
they are independent  of the gauge-fixing 
parameter,  and satisfy  naive  (ghost-free, QED-like) Ward identities
instead of the usual Slavnov-Taylor  identities.

For the specific problem of the NCR, by resorting to the PT   
one can construct  a  genuine,
target-independent form-factor endowed with  the crucial properties of
gauge-independence, gauge-invariance  and finiteness, for    arbitrary
values  of the    momentum transfer.  
The   essential ingredient   for
accomplishing this was the realization that,  when the target fermions
involved   are  considered   to   be    massless, all   gauge-dependent
contributions stemming from box  and vertex-like Feynman diagrams  are
effectively propagator-like, i.e.  have no dependence on the kinematic
properties or  quantum  numbers of  the  initial and final  states, as
boxes and vertices normally do. 
Thus,  all gauge dependence  stemming from vertices and boxes cancels
precisely against  the gauge-dependence stemming from the conventional
one-loop self-energies. As   a result, the remaining  gauge-independent
structures retain their initial kinematic identity; in particular one
can  speak in terms  of gauge-independent effective boxes, vertices, and
self-energies.  Thus,   once    the 
gauge-dependent pieces    have  been
extracted  from the box,   the remaining gauge-independent  ``pure'' box
should not be considered as a part of the resulting form-factor, which
should be entirely determined from the ``pure'' gauge-independent set of
vertex graphs. 

The aforementioned  propagator-like pieces are extracted 
by tracking down the action of 
the longitudinal momenta appearing in the $S$-matrix element of
$F\bar{F}\to \nu \bar{\nu}$ . Longitudinal momenta originate from the 
tree-level gauge-boson propagators and tri-linear gauge-boson vertices
appearing inside loops. In particular, in the $R_{\xi}$-scheme
the gauge-boson propagators have the general form  
\be
\Delta^{\mu\nu}_{\chic V} (k) = \Bigg [g^{\mu\nu}-
\frac{(1-\xi_{\chic V})k^{\mu}k^{\nu}}{ k^2 -\xi_{\chic V} 
M_{\chic V}^2}\Bigg ](k^2 -M_{\chic V}^2)^{-1}
\label{GenProp}
\ee
where $V=W,N$  with $N =  Z,\gamma$ and $M^2_{\gamma}=0$;  $k$ denotes
the virtual four-momentum circulating in the loop.  Clearly,  in the case of
$\Delta^{\mu\nu}_{\chic V} (k)$  the  longitudinal momenta are   those
proportional to $(1-\xi_{\chic V})$.
The longitudinal terms arising from the tri-linear vertex 
may be identified by splitting 
$\Gamma_{\alpha\mu\nu}(q,k,-k-q)$
appearing inside the one-loop diagrams 
into two parts ($q$ denotes the physical four-momentum entering
into the vertex):
\be
\Gamma_{\alpha\mu\nu}
= \Gamma_{\alpha\mu\nu}^{\chic F} + \Gamma_{\alpha\mu\nu}^{\chic P} \, ,
\label{decomp}
\ee 
with
\bea
\Gamma_{\alpha\mu\nu}^{\chic F}&=& 
(2k+q)_{\alpha} g_{\mu\nu} + 2q_{\nu}g_{\alpha\mu} 
- 2q_{\mu}g_{\alpha\nu} \, , \nonumber\\
\Gamma_{\alpha\mu\nu}^{\chic P} &=&
 -(k+q)_{\nu} g_{\alpha\mu} - k_{\mu}g_{\alpha\nu} \, .  
\label{GFGP}
\eea
The above decomposition assigns a special role 
to the $q$-leg,
and allows $\Gamma_{\alpha\mu\nu}^{\chic F}$ 
to satisfy the Ward identity
\be 
q^{\alpha} \Gamma_{\alpha\mu\nu}^{\chic F}= 
 (k+q)^2 g_{\mu\nu} - k^2  g_{\mu\nu}\, ,
\label{WI2B}
\ee
All aforementioned longitudinal momenta 
originating from $\Delta^{\mu\nu}_V (k)$ 
and $\Gamma_{\alpha\mu\nu}^{\chic P}$ 
trigger the following 
WI when contracted with the appropriate $\gamma$ matrix
appearing in the various elementary vertices. In the 
absence of fermion masses
\bea
\not\! k  P_{\chic L} &=& (\not\! k + \not\! p ) P_{\chic L} 
- P_{\chic R} \not\! p \, ,
\nonumber\\
&=& S_{\chic F'}^{-1}(\not\! k + \not\! p ) P_{\chic L} - 
P_{\chic R} S_{\chic F}^{-1}(\not\! p) 
\label{EWI}
\eea
where $P_{\chic R(L)} = [1  + (-) \gamma_5]/2$  is the  chirality projection
operator and   
$S_{\chic F}$ is the tree-level propagator of the fermion $F$;
$F'$ is the
isodoublet-partner of the external fermion $F$.
The result of this contraction is that 
the term in  Eq.(\ref{EWI}) proportional to  $S_{\chic F'}^{-1}$, i.e
the inverse of the internal fermion propagator      
gives rise to a self-energy-like term, 
whose coupling to the external fermions is proportional
to the vertex 
\be
\Gamma_{{\chic W} {\chic F} \bar{\chic F}}^{\mu} 
= -i\bigg(\frac{g_w }{2}\bigg) \gamma_\mu P_{\chic L}.
\ee
The effective vertex $\Gamma_{{\chic W} {\chic F} \bar{\chic F}}^{\mu}$ can be 
written as a linear combination of the two tree-level
vertices $\Gamma_{\gamma {\chic F} \bar{\chic F}}^{\mu}$ and
$\Gamma_{{\chic Z} {\chic F} \bar{\chic F}}^{\mu}$ 
as follows:
\be
\Gamma_{{\chic W} {\chic F} \bar{\chic F}}^{\mu}  
= \bigg(\frac{s_w}{2 T^{\chic F}_z}\bigg)
\Gamma_{\gamma {\chic F} \bar{\chic F}}^{\mu} - 
\bigg(\frac{c_w}{2 T^{\chic F}_z}\bigg)\Gamma_{{\chic Z} {\chic F}
\bar{\chic F}}^{\mu}
\label{SepVer}
\ee
In the above formulas
$Q_{\chic F}$ is the electric charge of the fermion $F$, and
$T^{\chic F}_z$ its $z$-component of the weak isospin,
with $c_w = \sqrt{1 - s^2_w} = M_{\chic W}/M_{\chic Z}$, $e=g_w s_w$. 
The identity of Eq.(\ref{SepVer})
allows to combine the propagator-like
parts  emerging from box-diagrams and vertex-diagrams
after the application of the WI in  Eq.(\ref{EWI})
with the conventional 
self-energy graphs $\Pi^{\gamma {\chic Z}}_{\mu\nu}$ and
$\Pi^{{\chic Z} {\chic Z}}_{\mu\nu}$ , by judiciously multiplying 
the former by 
$D_{\chic N} (q) D_{\chic N}^{-1}(q)$.

Finally, from the gauge-independent 
one-loop {\it proper} vertex 
$\widehat{\Gamma}^{\mu}_{A \nu_i \bar{\nu}_i}$ 
constructed using this method one extracts
the dimension-full electromagnetic
form-factor $\widehat{F}_{\nu_i}(q^2)$ as
$\widehat{\Gamma}^{\mu}_{A \nu_i \bar{\nu}_i}
= ie q^2  \widehat{F}_{\nu_i}(q^2) \gamma_{\mu}(1-\gamma_{5})$ .
The NCR, to be denoted by $\big <r^2_{\nu_i}\,  \big>$, is  
then defined as   
$\big <r^2_{\nu_i}\,  \big> = 6 \widehat{F}_{\nu_i}(0)$, and thus 
one obtains 
\be
\big <r^2_{\nu_i}\,  \big> =\, 
\frac{G_{\chic F}}{4\, {\sqrt 2 }\, \pi^2} 
\Bigg[3 
- 2\log \Bigg(\frac{m_{\chic i}^2}{M_{\chic W}^2} \Bigg) \Bigg]\, ,
\label{ncr}
\ee
where $i= e,\mu,\tau$, 
$m_i$ denotes the mass of the charged iso-doublet
partner of the neutrino under consideration, and $G_{\chic F}$
is the Fermi constant. 

\section{Measuring the probe-independent NCR}
After arriving at a physically menaingful definition for the 
NCR, the next crucial question is whether the NCR so defined 
constitutes a genuine physical observable. In the rest of this 
section we will briefly discuss 
the method proposed in  \cite{Bernabeu:2002nw}
for the extraction of the NCR from experiment. 

It is important to emphasize that  
measuring the entire process  
$f^{\pm}\nu \to f^{\pm} \nu$ does 
{\it not} constitute a measurement
of the NCR, because by 
 changing the target fermions $f^{\pm}$ one will 
generally change the answer, thus introducing a target-dependence
into a quantity which (supossedly) constitutes an
intrinsic property of the neutrino. 
Instead, what we want to measure is the target-independent
Standard Model NCR only, 
stripped of any target dependent contributions. 
Specifically, as mentioned above, the PT
rearrangement of the $S$-matrix
makes possible the definition of distinct, physically meaningful
sub-amplitudes, one of which,   
$\widehat{\Gamma}^{\mu}_{A \nu_i \bar{\nu_i}}$, 
is finite and directly
related to the NCR. 
However, the remaining sub-amplitudes, such as 
self-energy, vertex- and box-corrections,  
even though they 
do no enter into the definition of the
NCR, still contribute numerically to the entire $S$-matrix; 
in fact, some of them combine to form additional  
physical observables of interest, 
most notably the effective (running) 
electroweak charge and mixing angle.
Thus, in order to isolate the NCR, one must conceive of
a combination of experiments and kinematical conditions, 
such that 
all contributions not related to the NCR
will be eliminated.

Consider the elastic processes 
$ f(k_1) \nu(p_1)  \to f(k_2) \nu(p_2) $ and 
$f(k_1) \bar{\nu}(p_1)  \to f(k_2) \bar{\nu}(p_2) $,
where $f$ denotes an electrically charged 
fermion belonging to a different
iso-doublet than the neutrino $\nu$, in order to eliminate 
the diagrams mediated by a charged $W$-boson.
The Mandelstam variables are defined as
$s=(k_1+p_1)^2 = (k_2+p_2)^2$, 
$t= q^2 = (p_1-p_2)^2 = (k_1-k_2)^2$, 
$u = (k_1-p_2)^2 = (k_2-p_1)^2$, 
and $s+t+u=0$. 
In what follows we will restrict ourselves to the 
limit $t=q^2 \to 0$ of the above amplitudes,
assuming that all external (on-shell) fermions are massless.
As a result of this special kinematic situation we have the
following relations:
$p_1^2 = p_2^2 = k_1^2 = k_2^2 = p_1 \cdot p_2 = k_1 \cdot k_2 = 0$
and 
$p_1 \cdot k_1  = p_1 \cdot k_2 = p_2 \cdot k_1 = p_2 \cdot k_2 = s/2 $.
In the center-of-mass system we have that 
$t=-2 E_{\nu}E_{\nu}'(1-x)\leq 0 $, 
where $E_{\nu}$ and $E_{\nu}'$
are the energies of the neutrino before and after the
scattering, respectively, and 
$x \equiv \cos\theta_{cm}$, where
$\theta_{cm}$
is the scattering angle. Clearly, the condition $t=0$ 
corresponds to the exactly forward amplitude, 
with $\theta_{cm}=0$, \, $x=1$.

At tree-level the amplitude  $ f \nu  \to f \nu $ is 
mediated by an off-shell $Z$-boson, coupled to the fermions  
by means of the bare vertex 
$\Gamma_{Z {f} \bar{f}}^{\mu} = -i 
(g_w/ c_w)\, \gamma^{\mu}\, [ v_f + a_f \gamma_5]$
with 
$v_f = s^2_w Q_{f} - \frac{1}{2} T^f_z$  and 
$a_f=\frac{1}{2} T^f_z$.

At one-loop, the relevant  
contributions are determined  
through the PT 
rearrangement of the amplitude, giving rise to 
gauge-independent sub-amplitudes. 
In particular, the one-loop 
$AZ$ self-energy $\widehat{\Sigma}_{\chic{A}\chic{Z}}^{\mu\nu}(q^2)$
obtained is transverse, for {\it both} 
the fermionic and the bosonic contributions,
i.e. $\widehat{\Sigma}_{\chic{A}\chic{Z}}^{\mu\nu}(q^2)
= (q^2 \, g^{\mu\nu}  - q^{\mu} q^{\nu}) 
{\widehat{\Pi}}_{ \chic{A} {\chic Z}} (q^2)$.
Since the external currents are conserved, 
from the $ZZ$ self-energy 
$\widehat{\Sigma}_{\chic{Z}\chic{Z}}^{\mu\nu}(q^2)$
we keep only the part proportional to $g^{\mu\nu}$, 
whose dimension-full cofactor will be denoted 
by $\widehat{\Sigma}_{\chic{Z}\chic{Z}}(q^2)$.
Furthermore, as is well-known, 
the one-loop
vertex 
$\widehat\Gamma_{{\chic Z} {\chic F} \bar{\chic F}}^{\mu}(q,p_1,p_2)$, 
with $F = f$  or $F = \nu$, 
satisfies a QED-like 
Ward identity, relating it to the one-loop  
inverse fermion propagators $\widehat\Sigma_{\chic F}$,
i.e  
$ q_{\mu} 
\widehat\Gamma_{{\chic Z} {\chic F} \bar{\chic F}}^{\mu}(q,p_1,p_2)
= 
\widehat\Sigma_{\chic F} (p_1) - \widehat\Sigma_{\chic F} (p_2)$.
It is then easy to show that, in the limit of 
$q^2 \to 0$,  
$\widehat\Gamma_{{\chic Z} {\chic F} \bar{\chic F}}^{\mu} 
\sim q^2 \gamma^{\mu}(c_1 + c_2 \gamma_5)$; 
since it is multiplied by a
massive $Z$ boson propagator $(q^2 - M_{\chic Z})^{-1}$, its 
contribution to the amplitude vanishes when 
$q^2 \to 0$. This is to be contrasted with the 
$\widehat{\Gamma}^{\mu}_{A \nu_i \bar{\nu}_i}$,  
which is accompanied by a 
$(1/q^2)$ photon-propagator, thus giving rise 
to a contact interaction between the target-fermion and the neutrino,
described by the NCR. 
 
We next proceed to eliminate the target-dependent box-contributions;
to accomplish this we resort to the ``neutrino--anti-neutrino'' method.
The basic observation 
is that the tree-level amplitudes 
${\cal M}_{\nu f}^{(0)}$ 
as well as the
part of the one-loop amplitude ${\cal M}_{\nu f}^{(B)}$
consisting of 
the propagator and vertex corrections 
(the ``Born-improved'' amplitude),
are proportional to $[\bar{u}_{f}(k_2)\gamma_{\mu}( v_f + a_f \gamma_5 ) 
u_{f}(k_1)]
[\bar{v}(p_1)\gamma_{\mu} P_{\chic L} \, v(p_2)]$, 
and  
therefore transform differently than the boxes 
under the replacement 
$\nu \to \bar{\nu}$. 
In particular, 
the coupling of the 
$Z$ boson to a pair of on-shell anti-neutrinos 
may be written in terms of on-shell neutrinos
provided that
one changes the chirality projector from 
$P_{\chic L}$ to $P_{\chic R}$, 
and supplies a relative minus sign \cite{Sarantakos:1983bp}, i.e.
\be
\bar{v}(p_1)\gamma_{\mu} P_{\chic L} \, v(p_2)
= - \, 
\bar{u}(p_2)\gamma_{\mu} P_{\chic R} \, u(p_1)
\label{Gantinu}
\ee
Thus, under the above transformation, 
${\cal M}_{\nu f}^{(0)} + {\cal M}_{\nu f}^{(B)}$ reverse 
sign once, 
whereas the box contributions reverse sign twice.
These distinct transformation properties 
allow for the isolation of
the box contributions 
when judicious combinations of the forward differential cross-sections
$(d\sigma_{\nu f}/dx)_{x=1}$ and  
$(d\sigma_{\bar{\nu} f}/dx)_{x=1}$ 
are formed. In particular, 
$\sigma^{(+)}_{\nu f} \equiv 
(d\sigma_{\nu f}/dx)_{x=1}
+ (d\sigma_{\bar{\nu} f}/dx)_{x=1}$ does not contain 
boxes, 
whereas the conjugate combination 
$\sigma^{(-)}_{\nu f} \equiv 
(d\sigma_{\nu f}/dx)_{x=1}
- (d\sigma_{\bar{\nu} f}/dx)_{x=1}$  
isolates the contribution of the boxes.

Finally, a detailed analysis shows 
that in the kinematic limit we consider, 
the Bremsstrahlung contribution vanishes, 
due to a  a completely destructive interference 
between the two relevant  diagrams corresponding to the
processes $f A \nu (\bar{\nu}) \to f \nu (\bar{\nu})$ and 
$f \nu (\bar{\nu}) \to f A \nu (\bar{\nu})$. 
The absence of such corrections is consistent with the
fact that there are no infrared divergent contributions 
from the (vanishing) vertex
$\widehat\Gamma_{{\chic Z} {\chic F} \bar{\chic F}}^{\mu}$, 
to be cancelled against.  
 
$\sigma^{(+)}_{\nu f}$ receives contributions from the 
tree-level exchange of a $Z$-boson, the one-loop contributions
from the ultraviolet divergent quantities 
$\widehat{\Sigma}_{\chic{Z}\chic{Z} }(0)$ and  
${\widehat{\Pi}}^{ \chic{A}  {\chic  Z}} (0)$, 
and the (finite) NCR, coming from the proper vertex 
$\widehat{\Gamma}^{\mu}_{A \nu_i \bar{\nu}_i}$.
The first three contributions are universal, i.e. common to all 
neutrino species, whereas that of the proper vertex 
$\widehat{\Gamma}^{\mu}_{A \nu_i \bar{\nu}_i}$
is flavor-dependent. 

To proceed, the renormalization of 
$\widehat{\Sigma}_{\chic{Z}\chic{Z} }(0)$ and  
${\widehat{\Pi}}_{ \chic{A}  {\chic  Z}} (0)$ must be carried out.
It turns out that, by virtue of the Abelian-like Ward-identities 
enforced after the pinch technique rearrangement 
\cite{Cornwall:1982zr},
 the resulting expressions combine in such a way as to form manifestly 
renormalization-group invariant combinations 
\cite{Hagiwara:1994pw}.
In particular, after 
carrying out the standard re-diagonalization,
two such quantities 
may be constructed:
\bea
\bar{R}_{\chic{Z}}(q^2) &=& 
\frac{\alpha_w}{c_w^2}
\bigg[q^2 - M_\chic{Z}^2 +\Re e\,\{\widehat{\Sigma}_{\chic{Z}\chic{Z}}(q^2)\}
\bigg]^{-1}
\nonumber\\  
\bar{s}_w^{2}(q^2) &=&  s_w^{2}\Biggl(1 - \frac{c_w}{s_w}\, 
\Re e\,\{\widehat{\Pi}_{\chic{A}\chic{Z}}(q^2)\}\Biggr) \,.
\label{RW}
\eea
where $\alpha_w = g_w^2/4\pi$, 
and $ \Re e\,\{...\}$ denotes the real part. 
In addition to being renormalization-group invariant, 
both quantities defined in  Eq.(\ref{RW})
are universal (process-independent); 
$\bar{R}_{\chic{Z}}(q^2)$ corresponds to the $Z$-boson 
effective charge, while $\bar{s}_w^{2}(q^2)$  
corresponds to an effective mixing angle. 
We emphasize that 
the  renormalized  ${\widehat{\Pi}}_{ \chic{A}  {\chic  Z}} (0)$
{\it cannot}  form part of the NCR,  because 
it fails  to form a
renormalization-group invariant
quantity on its own.  
Instead,
${\widehat{\Pi}}_{\chic{A} {\chic Z}}  (0)$ must be combined with the
appropriate tree-level 
contribution (which  evidently  does not  enter into  the
definition of the NCR, since it is $Z$-mediated) 
in order to form the effective  
$\bar{s}_w^{2}(q^2)$ 
acting on the electron vertex,
whereas the finite 
NCR will be determined from the proper neutrino vertex only. 

After recasting $\sigma^{(+)}_{\nu f}$ 
in terms of manifestly 
renormalization-group invariant building blocks, one may 
fix $\nu = \nu_{\mu}$, 
and then consider three different choices for $f$: (i) 
right-handed electrons, $e_{\chic R}$; 
(ii) left-handed electrons, $e_{\chic L}$, and (iii) neutrinos, 
$\nu_{i}$ 
other than  $\nu_{\mu}$, i.e. $i=e,\tau$. 
Thus, we arrive at the system 
\bea
\sigma^{(+)}_{\nu_{\mu} \,\nu_i} &=& s \pi \bar{R}^2(0)
\nonumber\\
\sigma^{(+)}_{\nu_{\mu} \,e_{\chic R}}  &=& 
s \pi \bar{R}^2(0)\, \bar{s}_w^{4}(0) 
- 2 \lambda s_w^{2} \, 
\big< r^2_{\nu_{\mu}}\, \big>
\nonumber\\
\sigma^{(+)}_{\nu_{\mu} \,e_{\chic L}} &=& 
s \pi \bar{R}^2(0) \,
\bigg(\frac{1}{2} - \bar{s}_w^{2}(0)\bigg)^{2} \nonumber\\  
&& + \lambda (1-2 s_w^{2}) \, 
\big< r^2_{\nu_{\mu}}\, \big> 
\label{syst1}
\eea
where $\lambda \equiv (2\sqrt{2}/3) s \alpha \,G_{\chic F}$, 
$\alpha = e^2/4\pi$.
$\bar{R}^2(0)$, $\bar{s}_w^{2}(0)$, and  
$\big< r^2_{\nu_{\mu}}\, \big>$ are treated as three unknown 
quantities, to be determined from the above equations. 
The corresponding solutions are given by 
$\bar{s}_{w}^{2}(0) =  s_w^{2} \pm \Omega^{1/2}$ and 
\bea
\big< r^2_{\nu_{\mu}}\, \big> & = & \lambda^{-1} 
\Bigg[\bigg(s_w^{2}-\frac{1}{4} \pm \Omega^{1/2}\bigg)
\sigma^{(+)}_{\nu_{\mu} \,\nu_i}  
\nonumber\\
&&\qquad \qquad + \sigma^{(+)}_{\nu_{\mu} \,e_{\chic L}} - 
\sigma^{(+)}_{\nu_{\mu} \,e_{\chic R}}
\Bigg]\,
\label{sol2}
\eea
where the discriminant $\Omega$ is given by
\be
\Omega = (1- 2 s_w^{2}) \bigg(\frac{\sigma^{(+)}_{\nu_{\mu} \,e_{\chic R}}}
{\sigma^{(+)}_{\nu_{\mu} \,\nu_i}} -  
\frac{1}{2} s_w^{2}  
\bigg)
+  2 s_w^{2} 
\frac{\sigma^{(+)}_{\nu_{\mu} \,e_{\chic L}}}
{\sigma^{(+)}_{\nu_{\mu} \,\nu_i}}
\label{disc}
\ee
and must satisfy $\Omega > 0$. 
The actual sign in front of $\Omega$ may be chosen by requiring that 
it correctly accounts for the sign of the shift of $\bar{s}_{w}^{2}(0)$
with respect to $ s_w^{2}$ predicted by the theory \cite{Hagiwara:1994pw}. 
To extract the experimental values of the quantities 
$\bar{R}^2(0)$, $\bar{s}_w^{2}(0)$, and $\big< r^2_{\nu_{\mu}}\, \big>$,
one must substitute  
in Eq.(\ref{sol2}) and Eq.(\ref{disc}) the experimentally
measured values for the differential cross-sections 
$\sigma^{(+)}_{\nu_{\mu} \,e_{\chic R}}$, 
$\sigma^{(+)}_{\nu_{\mu} \,e_{\chic L}}$,
and $\sigma^{(+)}_{\nu_{\mu} \,\nu_i}$. 
This means that to solve the system one would 
have to carry out three different pairs of experiments. 

In summary, we have seen that the neutrino charge radius can be  
defined unambiguously by means of the pinch 
technique, and that its observability has been 
established.  

\medskip

{\it Acknowledgments}:
This work has been supported by the Grant AEN-99/0692 
of the Spanish CICYT. ~J.P. thanks the organizers of IMFP2002
for their hospitality at Jaca 
and the stimulating environment they have created.\\

\end{document}